\documentclass[a4paper,11pt]{article}
\usepackage{jinstpub} 

\title{\boldmath A new track finding algorithm based on a multi-dimensional extension of the Hough Transform}

\author{Luciano Ristori}
\affiliation{Fermilab, P.O.Box 500, Batavia, Illinois, 60510, USA}

\emailAdd{luciano@fnal.gov}

\abstract{
We introduce a new pattern recognition algorithm for track finding in High Energy Physics Experiments based on an extension of the Hough Transform to multiple dimensions.
A remarkable property of this algorithm is that the execution time is simply proportional to the total number of the hits to be processed, making it particularly attractive for high occupancy situations.
The algorithm needs to be {\em trained} using a sufficiently large set of simulated tracks. The same track finding algorithm can be used for very different detector geometries and only the set of simulated tracks used for training needs to be changed. The particular structure of the algorithm also lends itself naturally to parallel hardware implementations which, combined with its intrinsic flexibility, should provide a most
 powerful tool for triggering at future colliders.
}


\begin{document}
\maketitle
\flushbottom

\section{Goals of this work}
\label{}

For the future generation of particle colliders, efficient track finding algorithms will be crucial to process data both online and offline. All efforts to improve on the quality and performance of tracking algorithms are therefore amply motivated. We propose here to depart from the widely adopted class of Kalman filter based algorithms \cite{Kalman} and explore the potentiality  of a new class of algorithms loosely inspired by the Hough transform concept \cite{Hough}. One important goal is to try to reduce the time spent in making combinations of hits which plague the first stages of many currently adopted track reconstruction algorithms and cause the execution time to grow with some power of the hit density. This is problematic when the number of hits is very high.
Ideally we would like an algorithm whose execution time grows linearly with the number of hits. 
We also would like the capability of easily adapting to different geometries. For example we would be able to provide a sample of simulated tracks  with the corresponding hits generated in a detector of arbitrary geometry and the code would {\em learn} how to perform pattern recognition and extract track parameters without the need to change the algorithm.
We also would like to be able to include the time of arrival of the hit, if available, as an additional coordinate to be treated in the same way as a spatial coordinate, both in the pattern recognition stage (to find the track and discriminate it from background) and in track fitting stage (to extract track parameters, possibly including the mass of the particle).

\section{Track finding: a simple example}
\label{}

\begin{figure}[htbp]
\begin{center}
\vspace{12pt}
\includegraphics[scale=0.2]{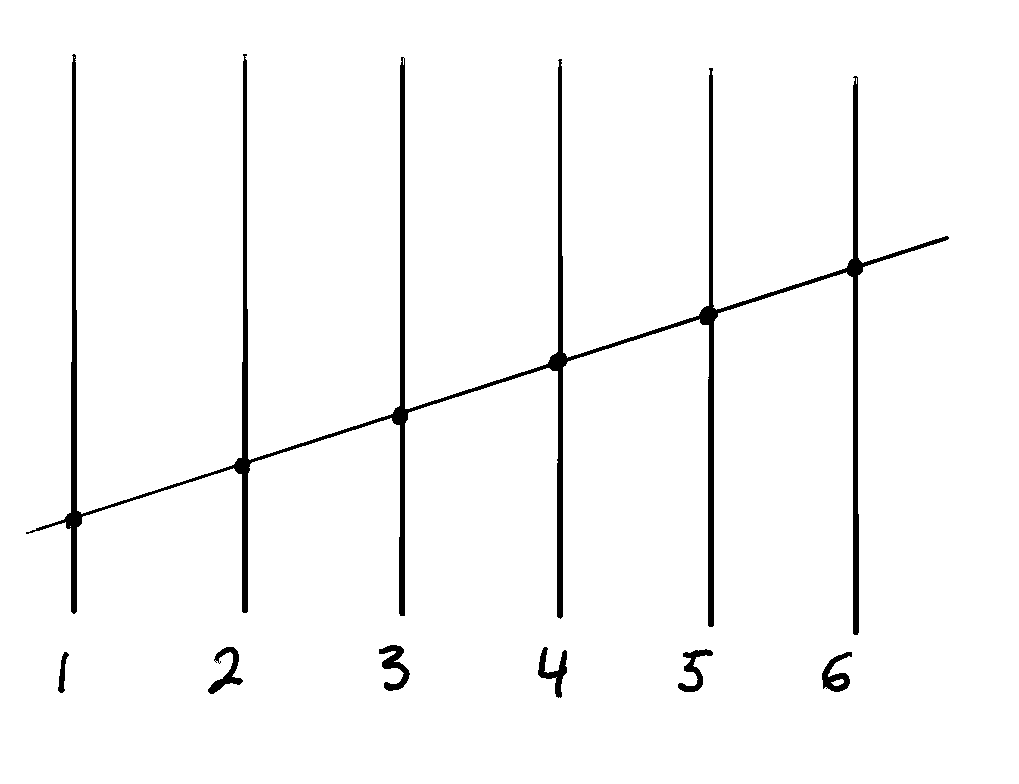} 
\vspace{12pt}
\caption{A simple example of a straight track going through six parallel detector planes measuring a single coordinate.
The track can be identified with two parameters $(a, b)$ and each hit with a single number $x$.
}
\label{simple}
\end{center}
\end{figure}

One simple possible application of the Hough Transform is when the track parameter space is 2-dimensional and the hit coordinates are one dimensional, for example when straight tracks traverse a series of parallel flat detectors measuring a single coordinate (fig.\ref{simple}).

Straight tracks can be described by two parameters and the intersection of the track with each detector can be described by a single coordinate $x$. If we call the two parameters of the track $a$ and $b$, we can identify each possible track with a point in a 2-dim plane whose axes are labeled $a$ and $b$.

\begin{figure}[htbp]
\begin{center}
\vspace{12pt}
\includegraphics[scale=0.35]{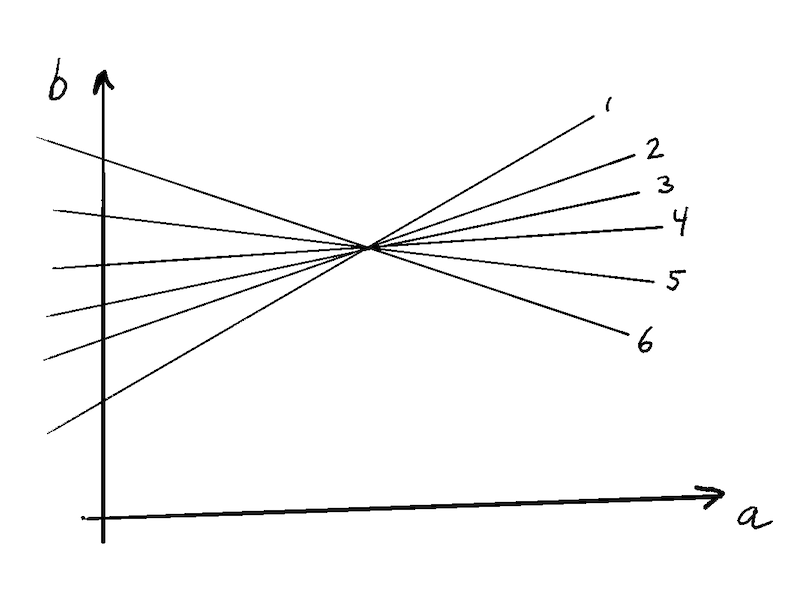} 
\vspace{12pt}
\caption{Representation of a two dimensional straight track parameter space on a two dimensional $(a, b)$ plane.
All the tracks going through a single point on a specific detector form a straight line in the $(a, b)$ plane.
All these lines cross in correspondence of the $(a, b)$ parameters of the track that generated all points.
}
\label{ab-plane}
\end{center}
\end{figure}

Given the coordinate of a single hit on a particular detector plane, we can draw all the points that correspond to all the possible tracks that would go through that hit. All those points will form a line. Hits on different detector planes will form different lines. All the lines will intersect in a single point corresponding to the parameters of the track that originated all the hits (fig.\ref{ab-plane}).

\begin{figure}[htbp]
\begin{center}
\vspace{12pt}
\includegraphics[scale=0.3]{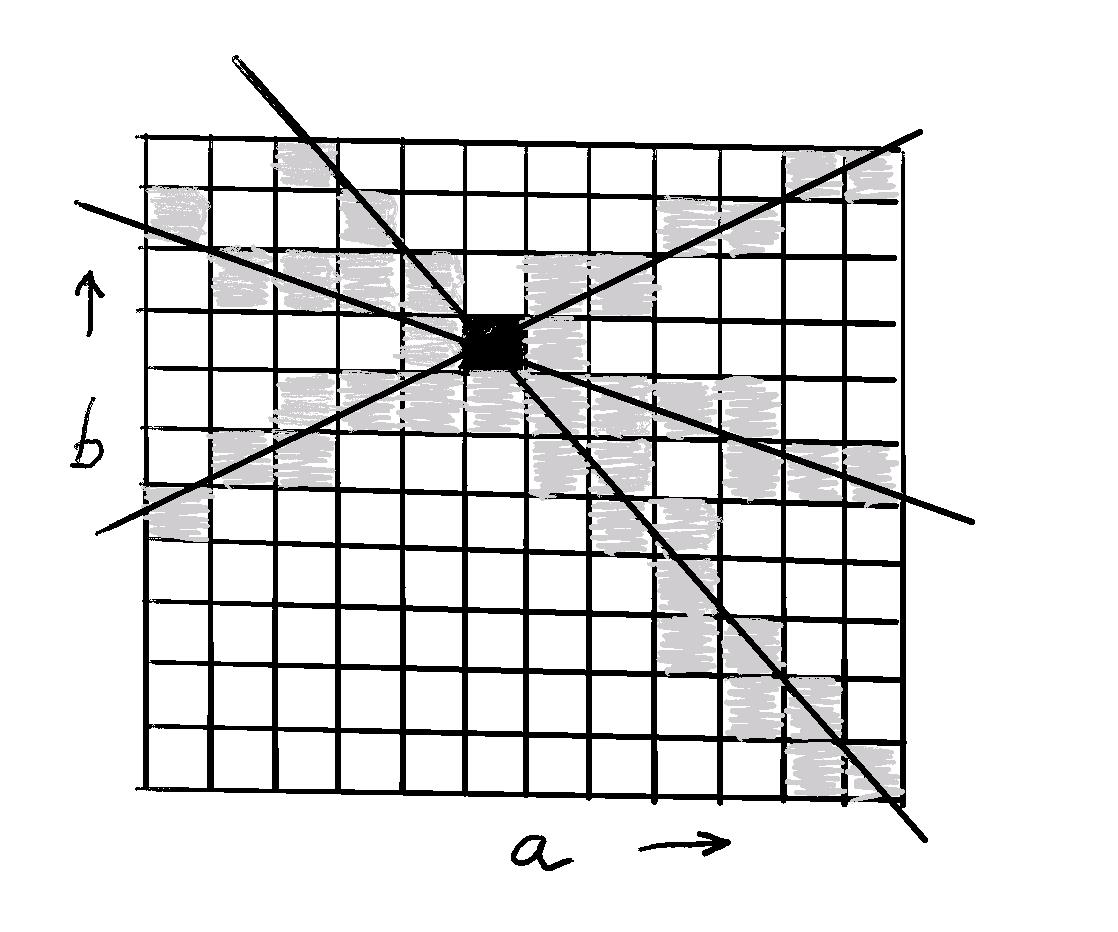} 
\vspace{12pt}
\caption{Here is a conceptual representation of a Hough Transform Array for two-parameter straight tracks $(a, b)$.
It can be thought as a quantized version of fig.\ref{ab-plane}.
Each hit is used to mark all the cells containing parameters of tracks compatible with generating that hit. After all the hits have been examined, the cell marked by the most hits is the one most likely to contain the best track candidate.
}
\label{2dimHTA}
\end{center}
\end{figure}

Following this observation, to write a simple pattern recognition algorithm based on the Hough Transform concept, one can take the $a-b$ plane and subdivide it into a number of small cells, then,  for each hit,  one would mark all the cells which contain at least one value of the parameters of a track compatible with producing that particular hit and we would attach that hit to all those cells.  In the end, the cell where all the hits, or the maximum number of hits are attached, is the one most likely to contain the correct value of the track parameters.
We refer to the ensemble of all the cells in the track parameter space as the Hough Transform Array (HTA)
(fig.\ref{2dimHTA}).

\section{Training}
\label{training}

From the computational point of view, the most demanding task during pattern recognition is finding out, given one hit, to which cells in the HTA it should be attached. If we have access to a simulator that, given the parameters of a track, calculates the coordinates of the hits that track produces on each detector layer, then this problem can be solved by a learning process during a training phase. The learning process consists in generating, for each cell in the HTA, a sufficient number of tracks, uniformly distributed inside the cell, computing the value of the coordinates of each hit, taking the maximum and minimum values of $x$ for each detector layer and storing them in a database.  
Extreme values must be stored permanently, for each HTA cell and each detector layer.

To know which cells a given hit needs to be attached to, one needs to check if its coordinate $x$ falls within the minimum and maximum values obtained in the training phase for that cell and for that detector layer.
The pattern recognition process amounts to going through all the hits and attaching each one of them to all the cells it needs to be attached to and then selecting those cells with a sufficient number of hits attached and declaring them a {\em candidate}.

{\em Candidates} will normally be delivered to a fitting stage to extract track parameters and to resolve possible residual ambiguities caused by multiple hits in the same detector layer for a single candidate.

In the presence of multiple tracks and/or background hits, cell granularity must be sufficiently thin to ensure that, in the majority of cases, candidates will have only one single hit in each detector layer.

\section{Extension to multiple dimensions}
\label{}

\begin{figure}[htbp]
\begin{center}
\vspace{12pt}
\includegraphics[scale=0.4]{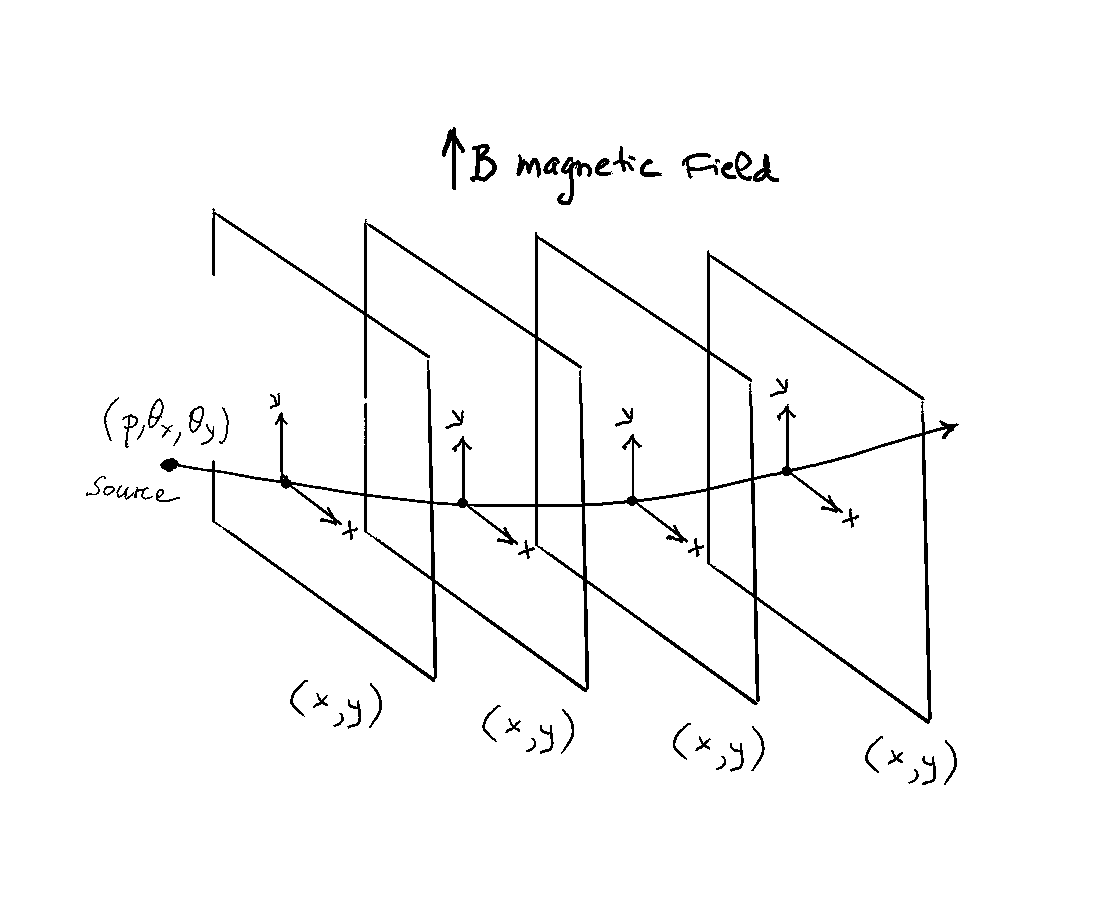} 
\vspace{12pt}
\caption{A multi-dimensional case example: charged tracks in a magnetic field coming from a point-like source. 
Each detector plane measures two orthogonal coordinates $(x, y)$. Each track is identified by three parameters $(p, \theta_x, \theta_y)$.
In this case the HTA array needs to be three-dimensional and both $x$ and $y$ limits need to be found during training for each array cell and each detector layer.
}
\label{MultiDim}
\end{center}
\end{figure}

The concept of Hough Transform can be extended to multiple dimensions both in track parameter space and in hit coordinate space. 
We will then talk of a Multi Dimensional Hough Transform (MDHT).
For example we can consider charged tracks in a magnetic field coming from a point-like source (3 parameters) going through a series of parallel plane detectors, each one measuring two coordinates $(x, y)$
(fig.\ref{MultiDim}).

In this case the cells in parameter space will be arranged in a three-dimensional array.  The learning process will be the same as in the previous example, the only difference being that now one needs to find the minimum and maximum values of both coordinates $x$ and $y$ and during pattern recognition one will need to check that both coordinates of the hit under consideration fall within the limits in order to attach it to a given cell.

It is important to realize that this whole process does not assume any specific geometric quality of the tracks or the detectors.  The only requirement is that,  given the parameters of the track, one can obtain, by simulation, the coordinates of the hits produced by that track in all detector layers.

\section{The Hough Transform Array (HTA)}
\label{}

Given the span of track parameters that one needs to cover, one must decide the number of bins into which to subdivide each one of them and thus  the total number of cells in which the overall parameter space will be subdivided. 
In this way we construct the Hough Transform Array (HTA): a quantized version of the track parameter space with one array index running for each different track parameter.
All the results of the training phase for a particular detector configuration are stored in the Hough Transform Array (HTA). 
For each cell in parameter space a list is created which includes all detector layers which are possibly hit by tracks with parameters inside that cell.
Each layer in the list contains the minimum and the maximum values of all the coordinates of the hits generated on that layer by the tracks from that cell.
All these data are stored permanently at the end of the training phase and retrieved any time one needs to perform pattern recognition.
During pattern recognition, each hit is compared with all the cells to find the appropriate detector layer and the corresponding minimum and maximum values of the coordinates found during training.
In case all the coordinates of the hit match with the limits, that hit is attached to that cell. 
Each hit is attached to all the cells it matches to.
As already explained for the simple 2-dim case (Sec. \ref{training}), a good candidate is declared when a sufficient number of hits is attached to a single cell.

The size of the cells of the Hough Transform Array is critical and should be taylored to the needs of each particular situation.
Smaller cells will reduce the number of track candidates contaminated by noise hits or by hits belonging to different tracks, while it will increase the amount of storage needed to hold the contents of the whole array and  the number of cells to be examined for each hit during the pattern recognition process and, consequently, the processing time.
On the other hand, larger cells, despite saving space, will increase the probability of additional hits from noise or adjacent tracks to be included in track candidates 
where they do not belong
and require multiple fits to decide which hit combination is the one most likely to be the correct one. It follows that, to minimize the execution time, one will have to search for a {\em sweet spot} in the number of bins which will depend on the particular situation, including hit density and coordinate measurement precision and will typically have to be studied using detailed simulation. Simulation will also be used to determine other tunable parameters like the minimum number of hits required to fit a track candidate. 
Compromises  will need to be reached between execution time, track efficiency and fake track rate.
Reducing the fake rate will favor an increase of the number of bins in the HTA array at the expense of execution time. Purity will also benefit from increasing the minimum number of hits required to fit a track, possibly eroding efficiency.

\section{Noise}
\label{}

Problems with pattern recognition may arise when multiple tracks traverse the detector at the same time or when {\em noise} hits are present. By {\em noise hits} we mean hits generated by tracks not related to the physics process we are studying, or by other physics processes like, for example, thermal noise or radioactivity.  In some circumstances hit density on the face of detectors may become sufficiently high to create confusion and making it difficult to identify which hits belong to which track

With this type of algorithm, in case of high hit density, we may find that some candidates come with more than one hit in some detector layers. In that case we will have different possible hit combinations to consider for a fit and will have to find a way to solve the ambiguities. For example we can fit all possible combinations and choose one using some {\em goodness-of-fit} criteria. Solving these ambiguities requires computing time and is therefore desirable to minimize them. A way to minimize multiple hits in the same cell and same detector layer is to have smaller cells in parameter space. That means, of course, more cells for a given track parameter span.

\section{Execution time}
\label{}

An important quality of this algorithm is that the execution time is approximately proportional to the total number of hits to be processed. This is in contrast with many algorithms that are based on constructing hit combinations using a trial-and-error strategy, where the execution time may typically grow with some power of the number of hits. A linear behavior of the execution time as a function of the number of hits is a very desirable feature to minimize possible problems at high hit densities.

\section{Array scan strategies}
\label{}

In the algorithm as described above, each hit is compared with all the cells in the array. Actually this is not a very efficient way to proceed. It can be easily seen that the large majority of these comparisons will be unsuccessful, resulting in a large waist of time. Instead, in most cases, just by knowing the coordinates of a hit, we can set limits to the range of the parameters of the parent track and, consequently, limit the number of cells we need to examine in detail, with a possible significant gain in execution time. In the  example given above in fig.\ref{MultiDim}, with tracks coming from a point-like source, it should be rather obvious how we can use the knowledge of the y coordinate of a hit to limit the $\theta_y$ search range.
This strategy can also be studied and optimized by simulation. Here the compromise to be researched is between execution time and efficiency because the exclusion of some cells from the search may cause some loss of hits and, ultimately, some track inefficiency.
Here some sufficiently small loss in efficiency might be justified by a significant gain in execution time.

\section{Possible hardware implementation}
\label{}

For applications where execution speed needs to be pushed to the limits (e.g. trigger applications in particle physics experiments), this MDHT algorithm lends itself to be implemented in hardware (FPGA, ASIC, ...) very naturally. In fact, each cell of the Hough Transform Array can be associated to a physical device performing the simple operation of comparing one hit with the maximum and minimum coordinate limits of that particular cell and storing that hit if and only if all the coordinates are within the limits.
The hits can be broadcast to all the cells at the same time and the cells can all work in parallel forming all track candidates in the time it takes to broadcast all the hits.
This massively parallel architecture resembles very closely that of the Associative Memory (AM) \cite{AssMem1}\cite{AssMem2}\cite{AssMem3} the main difference being that, in the AM, hit coordinates are required to match  specific stored values, where here, instead, we store minimum and maximum values and each coordinate is required to be within an interval. 
Present technology allows the implementation of this kind of operations with speeds in the GHz range, so this opens up the possibility of solving track finding for events with millions of hits in the millisecond time range and, given the intrinsic parallelism of the algorithm, the implementation of all the cells needed for a particular application can be distributed over a large number of identical building blocks (ASICs, FPGAs, GPUs, CPUs) making it, virtually, an infinitely scalable architecture.

\acknowledgments

I am very thankful to Massimo Casarsa and Sergo Jindariani for useful suggestions, for reviewing the manuscript and for helping in the submission process of this paper.




\end{document}